\RequirePackage{fix-cm}
\documentclass[twocolumn]{svjour3}                     

\smartqed  
\usepackage{graphicx}

\usepackage[hyphens]{url}
\usepackage[pdftex]{thumbpdf}
\usepackage[
    colorlinks,bookmarks=true,bookmarksnumbered=true,hypertexnames=false,citecolor=blue,urlcolor=blue,
    breaklinks=true,linkbordercolor={1 1 1},pdfborder={0 0 0}]{hyperref}
\hypersetup{
    pdfauthor={DMF},
    pdftitle={Artefacts in Software Engineering: A Fundamental Positioning},
    pdfkeywords={software engineering fundamentals, artefact orientation, position statement},
    colorlinks=true
}

\journalname{International Journal on Software and Systems Modeling}
\begin{document}

\title{Artefacts in Software Engineering: A Fundamental Positioning}


\author{Daniel M\'endez Fern\'andez\and
        	   Wolfgang B\"ohm \and
	   Andreas Vogelsang \and
	   Jakob Mund \and
	   Manfred Broy \and
	   Marco Kuhrmann \and
	   Thorsten Weyer
}


\institute{
	Daniel M\'endez Fern\'andez, Wolfgang B\"ohm, Manfred Broy\at
	Technical University of Munich\\
	Boltzmannstr. 3, 85748 Garching, Germany\\
        \email{Daniel.Mendez@tum.de, boehmw@in.tum.de, broy@in.tum.de}           
        \and
        Andreas Vogelsang\at
        Technical University of Berlin\\
	Ernst-Reuter-Platz 7, 10587 Berlin, Germany\\
        \email{andreas.vogelsang@tu-berlin.de}
        \and
        Jakob Mund\at
        Tableau Germany GmbH\\
        Maximilianstr. 35, 80539 Munich, Germany\\
        \email{JMund@Tableau.com}
	\and
	Marco Kuhrmann\at
	Clausthal University of Technology\\
	Wallstr. 6, 38640 Goslar, Germany \\
	\email{kuhrmann@acm.org}
	\and
	Thorsten Weyer\at
	University of Duisburg-Essen\\
	Gerlingstra{\ss}e 16, 45127 Essen, Germany\\
	\email{thorsten.weyer@paluno.uni-due.de}     
}

\date{Received: date / Accepted: date}

\maketitle

\begin{abstract}
	Artefacts play a vital role in software and systems development processes. Other terms like documents, deliverables, or work products are widely used in software development communities instead of the term artefact. In the following, we use the term `artefact' including all these other terms. Despite its relevance, the exact denotation of the term `artefact' is still not clear due to a variety of different understandings of the term and to a careless negligent usage. This often leads to approaches being grounded in a fuzzy, unclear understanding of the essential concepts involved. In fact, there does not exist a common terminology. Therefore, it is our goal that the term artefact be standardised so that researchers and practitioners have a common understanding for discussions and contributions. In this position paper, we provide a positioning and critical reflection upon the notion of artefact in software engineering at different levels of perception and how these relate to each other. We further contribute a meta model that provides a description of an artefact that is independent from any underlying process model. This meta model defines artefacts at three levels. Abstraction and refinement relations between these levels allow correlating artefacts to each other and defining the notion of related, refined, and equivalent artefacts. Our contribution shall foster the long overdue and too often underestimated terminological discussion on what artefacts are to provide a common ground with clearer concepts and principles for future software engineering contributions, such as the design of artefact-oriented development processes and tools.

\keywords{Software Engineering Artefacts \and Metamodelling \and Propaedeutics \and Syntax of artefacts \and Semantics of artefacts \and Equivalence of artefacts}
\end{abstract}

\section{Introduction}
\label{sec:Introduction}
Software development projects, especially when being complex, embody a multitude of artefacts that are developed, used, maintained, evolved, and which depend on each other. The final software product as such constitutes an artefact as well as the models of the software product, its specifications, safety and security certificates, project management plans, user manuals, the code, and many more. In software projects, these artefacts usually manifest themselves in the form of printed or electronic documents in the sense of coarse-grained structured collections of information serving distinct purposes. Even the binaries that make up a software product are just a complex conglomerate of files carefully put together such that they can be executed by computers and, thus, are artefacts. There is another dimension to this aspect: as artefacts can appear in different representations, for instance as files on a computer, as elements of data bases, or as printed documents, the question arises if all these are to be seen as one artefact or if they constitute different artefacts, and if so, how we can describe their relationship. Obviously, artefacts play a vital role in software and system development. Nevertheless, what an artefact actually is and how it is structured, is not obvious and not uniquely defined. 

In fact, software development process models make excessive use of artefacts to describe the products and documentation being created during the process. Just following a process and neglecting a clear notion of artefacts and their dependencies in purely process-driven approaches leads to problems and difficulties that have been recognised for years~\cite{Braun+05}, and the consequences can be observed in everyday industrial project settings~\cite{MWLBC10} including, for example, unclear roles and responsibilities or inconsistent and even incompatible specification documents provided by different teams. 

This brings us to a development paradigm known as ``artefact orientation''. Any artefact-oriented approach is rooted in a blueprint of the anticipated results and documents (and their structure, contents, and dependencies) instead of a blueprint of interconnected activities and method descriptions. The actual software development abstracts from the actual processes for creating those results. Since software systems as well as their descriptions and documentation can reach a considerable size and diversity, it is essential to adequately structure the set of artefacts produced in a development process via an artefact model~\cite{Boehm+13,Butting+17}. Those artefact models abstract from specific development processes while capturing the structure and content of the work products and, moreover, the established (modelling) concepts applied to the respective application domain. Artefact models provide a basis for a process-agnostic flexible backbone for project operation in which the team agrees on the content and the structure of the artefacts to be created until a specific point in time rather than a strict process and their synchronisation~\cite{KMG13}. At the same time, artefact models leave open the way of creating the results yet having a clear notion of responsibilities assigned to each artefact to be created. For software processes, artefact orientation has become an indispensable tool to support flexibility in the project lifecycle, precision in the results, and towards standardisation of terminology within and across projects. 

\paragraph{Problem Statement}
So far, our community still struggles with a common precise understanding and agreement about what a software engineering artefact exactly is and which terminology to use. Without such an understanding, researchers and practitioners will continue to discuss or even provide conceptual and formal contributions grounded in a fuzzy, unclear, or even contradictory understanding of the essential concepts involved. 

\paragraph{Contribution}
In this position paper, we suggest a foundation of the term artefact that is independent from the underlying process and provide a conceptual model that captures the essence of artefacts, i.e. the artefact structure and its semantics, using a descriptive approach. By taking this descriptive view, we characterise artefacts independently of the generating development processes. We provide a basis for further applications of the concept of an artefact such as a formal approach to describe semantic relationships between artefacts.

We also address the problem of artefacts that contain identical semantic information, by introducing the notion of \emph{equivalence} with respect to artefacts, which cannot only occur through different representations but also by the use of different description techniques for the same content. This notion of equivalence helps understanding the relations between software engineering artefacts. From these relations, we can derive, for instance, consistency rules and verification obligations.

\paragraph{Outline}
The remainder of the paper is structured as follows: In Section~\ref{sec:ArtAndLevOfPerception}, we present our characterisation of artefacts and explain the idea using a simple example in Section~\ref{sec:Example}. Section~\ref{sec:EquivArte} addresses the question of equivalence of artefacts. In Section~\ref{sec:RelatedWork}, we discuss related work. Finally, we conclude by discussing impacts and implications of our work in Section~\ref{sec:Conclusion}.

\section{Artefacts and Levels of Perception}
\label{sec:ArtAndLevOfPerception}
As discussed above, in a process-driven setting, the role of artefacts in the process and the activities used to create or modify these artefacts influence the way we see and understand them. We decide upon their content and structure by considering questions like which methods and description techniques we may or may not use for artefacts. This opens the doors for different and sometimes competing perceptions of artefacts and artefact models resulting in a blurry and overloaded terminology. 

We take a stakeholder-centric view: To extract the meaning of an artefact from its current representation, a stakeholder\footnote{In this context, we use the term ``stakeholder'' to denote humans building artefacts as well as humans making use of artefacts.} typically applies different levels of perception. In the following, we introduce these levels and the corresponding processing steps: The first thing that a stakeholder perceives when processing artefacts is their \emph{physical representation}. The physical representation of an artefact is, for example, a piece of paper, texts in a text processing system, a file in a file system, entries in a tool, or a record in a database. To use an artefact, a stakeholder will first parse the physical representation to extract the \emph{syntactic information} (structure) of the artefact. In this step, the stakeholder abstracts from all information that is only relevant for the physical representation, but irrelevant for the underlying syntactic structure. An example for such irrelevant information is whether the sections of one document are contained in one file or distributed across several files. Typically, a stakeholder differentiates between structures used to arrange information in containers (e.g., sections in a document), and structures used to express associated chunks of information (e.g., classes in a class diagram). Stakeholders typically know several structure types as part of their individual knowledge. Their knowledge influences how stakeholders perceive artefacts. While one stakeholder may recognise a physical representation as a class diagram, another one might only identify boxes, lines, and text. 

In a second step, the stakeholder interprets the syntactic representation to identify the underlying \emph{semantic content}, i.e. the ``meaning'' of the artefact. In this interpretation step, the stakeholder associates the structure with statements about the real world (and its interpretation), statements from a fixed semantic theory, or a mixture of both. Again, this is strongly influenced by the stakeholder's individual knowledge, experiences, abilities, and expectations, i.e. the level of expertise in a specific semantic domain. This context-sensitive meaning is also referred to as \emph{pragmatism} (see also~\cite{Lindland+94}).

Remarkably, this analysis of processing artefacts holds for humans and machines alike. If an artefact is processed by a machine, this machine needs to parse the physical representation to extract an underlying structure and then interpret the parsed structure to exhibit a specific behaviour that we consider as the artefact's content. Artefacts that are specifically made for (non-human) applications usually make the underlying structures explicit (e.g., by a file name extension). Moreover, even the use of a specific interpreter can be specified (e.g., by using a virtual machine). 

\begin{figure}[htbp]
\sidecaption
\includegraphics[width=0.5\textwidth]{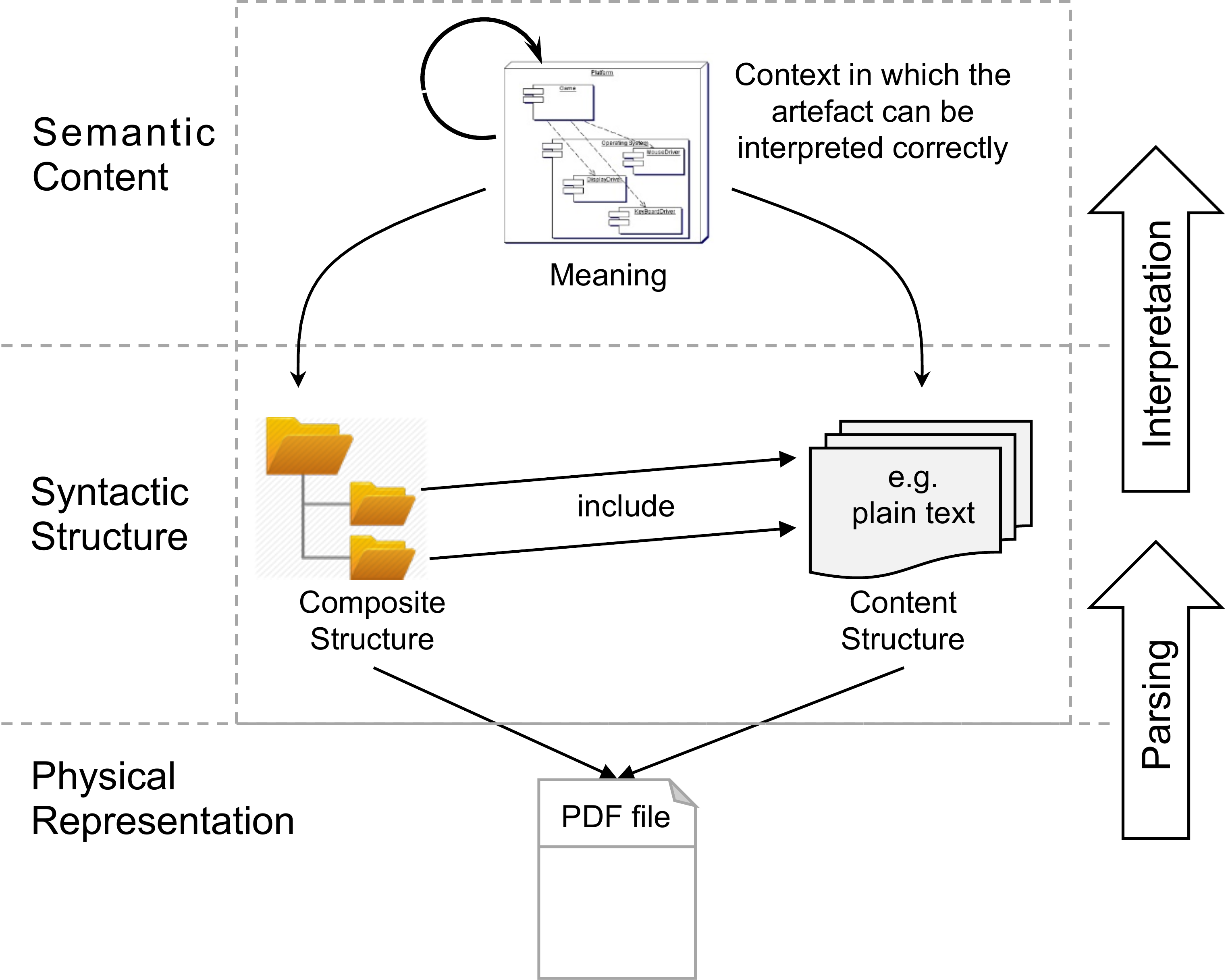}
\caption{The different levels of perception of an artefact.}
\label{fig:LevelsPerfecption}
\end{figure}

Based on this consideration, we characterise an artefact by the three levels of perception and the two processing steps mentioned above as visualised in Fig.~\ref{fig:LevelsPerfecption}.

\begin{description}[Level 3]
	\item[\textbf{Level 1 -- Semantic Content}] The content represents the meaning of an artefact. The content is always interpreted in the context of the individual knowledge of the stakeholder or the interpreter of the machine (the pragmatism). This individual knowledge contains a set of semantic theories and experiences about the real world. If this individual knowledge includes a semantic theory in which the content can be interpreted (e.g., via ontologies or mathematical models) the meaning is defined within the semantic theory -- the meaning is explicit. Otherwise, readers interpret the content in their understanding of the real world -- the meaning remains implicit. The content is usually interpreted partially by a semantic theory and partially by the real-world understanding of the reader.
	\item[\textbf{Level 2 -- Syntactic Structure}] The structure of an artefact describes how its content is expressed syntactically. The structure often includes a composition of several sub-artefacts. A \emph{composite structure} captures this composition (e.g., as in a table of contents of an artefact). The leaves of this composite structure define the lowest granularity of (sub-)artefacts. For the structure within these leaf artefacts, we distinguish between different \emph{content structures} used to represent the content (e.g., natural language, formal language like formulas or programs, figures, tables, diagrams, or mixtures thereof). Different types of artefact structures can be described by grammars or meta models.
	\item[\textbf{Level 3 -- Physical Representation}] An artefact has a physical representation. If a specific tool or representation language is used (such as Microsoft Word or \LaTeX) and the artefact is represented in that specific context of the format, then this may introduce new information at the physical representation level only (e.g., by adding \LaTeX\ environments) or even additional files (e.g., by generating log files during the typesetting process). This is, however, without specific relevance for the semantic content of the artefact. The physical representation therefore induces, to a certain extent, also some syntactic elements, which are not covered in the structure, because they are without any semantic relevance.
\end{description}
The processing steps for artefacts discussed above connect the three levels:
\begin{description}[Processing]
	\item[\textbf{Processing Step 1 -- Parsing}] Parsing describes the process of identifying a structure from the physical representation. The parsing process connects the physical representation level with the structure. The outcome of the parsing process is the composite structure of the artefact and the representation of the information chunks. The parsing is based on individual knowledge of the stakeholder. While one stakeholder may recognise a physical representation as a class diagram, another might only recognise boxes, arrows, and text.
	\item[\textbf{Processing Step 2 -- Interpretation}] Interpretation is the process of extracting the content (i.e. the meaning) from the structure. It connects the structure with the content. This results in an understanding of the meaning in the context of the stakeholder's knowledge. Which interpreter is selected and applied depends on that stakeholder's knowledge. The structure influences the selection of a specific interpreter, e.g., for a certain semantic theory. If the stakeholder is familiar with this semantic theory, this theory can be used to extract the meaning of the artefact. Otherwise, the stakeholder may interpret the content in the context of the individual real-world understanding.
\end{description}

\section{Example for Artefacts and Levels of Perception}
\label{sec:Example}
\begin{figure}[htbp]
\centering
\includegraphics[width=0.5\textwidth]{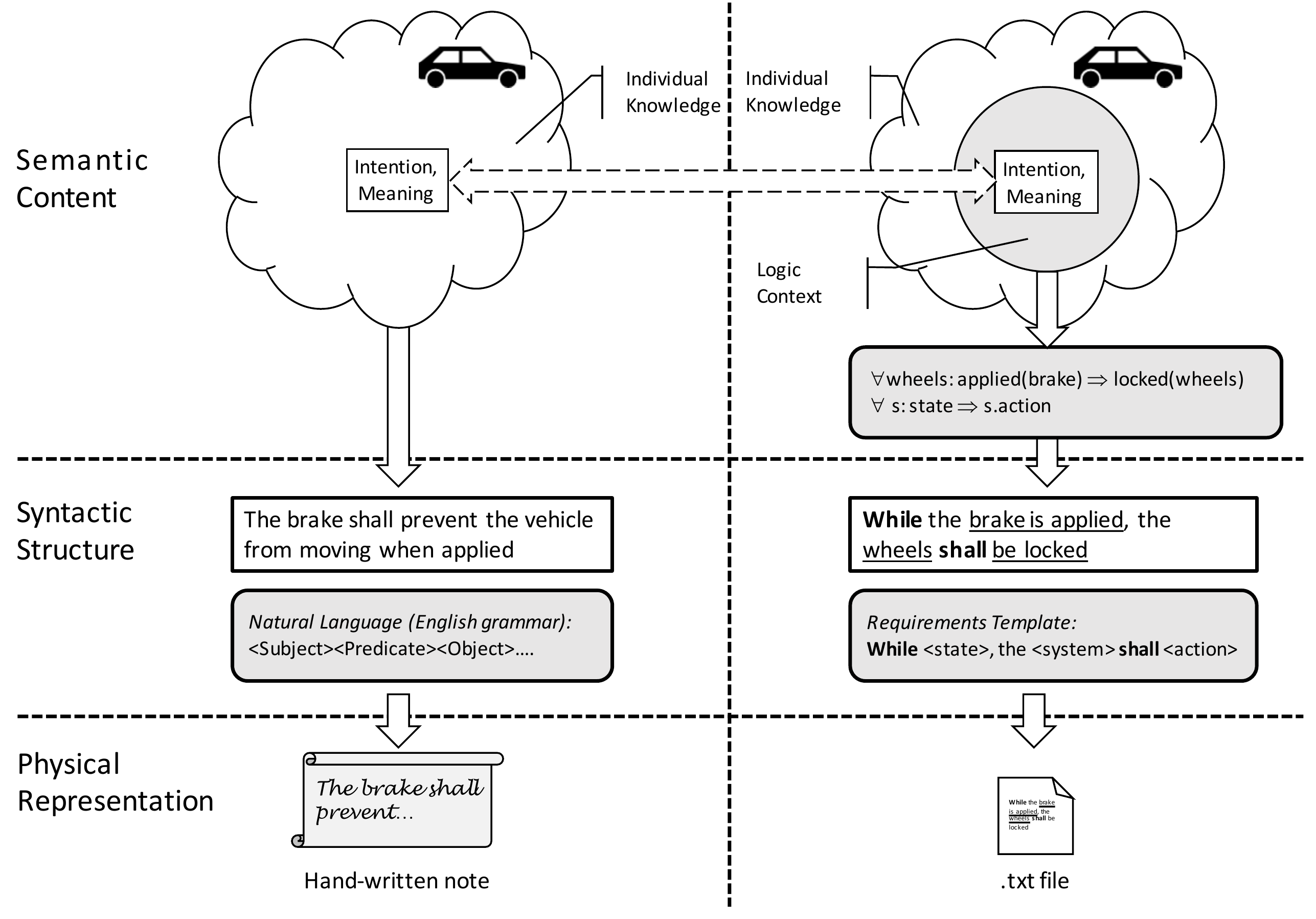}
\caption{Example of two artefacts with the same semantic intention.}
\label{fig:Example}
\end{figure}

To illustrate this characterisation and its usefulness, we consider a small example where a product manager thinks of a new requirement for a vehicle braking system and writes the following note on a piece of paper: \emph{``The brake shall prevent the vehicle from moving when applied''}. By taking this note, the product manager creates an artefact. Let us consider this artefact from the three levels introduced above. The physical representation of the artefact is the piece of paper on which the requirement is written down (see Fig.~\ref{fig:Example}, left side). From the syntactic point of view, the product manager uses natural language as \emph{content structure} of the requirement. In this case, the English orthography and grammar defines the syntactic rules of the content's structure. There is no additional \emph{composite structure} in this simple example since there is only one requirement. In the centre of the semantic view is the \emph{meaning}, the artefact's semantics that the product manager intends to express. This \emph{meaning} emerges when the product manager or some other stakeholder working with the artefact interprets the \emph{content} in the context of her individual knowledge. A person with a different individual knowledge might associate the same syntactic appearance with a different meaning.

Consider now a requirements engineer who is responsible for collecting and managing all requirements for a vehicle. The requirements engineer talks to the product manager about the requirement and about its meaning and documents it in a file. However, the requirements engineer might first rewrite the requirement to express it using a requirements template (\emph{constrained natural language}\/). By doing so, the requirements engineer also creates an artefact, which, however, differs from the first with respect to the three levels of perception. Please recall that the semantics (meaning) of both artefacts is supposed to be the same. The physical representation of the new artefact is given by a file that contains the requirement as text and that is stored in a file system (see Fig.~\ref{fig:Example}, right side). From the syntactic point of view, the requirements engineer uses a requirements template as \emph{content structure} of the artefact. The requirements template defines the constructs and rules of the appearance. More specifically, the template provides a meta model, even if kept implicit, that defines syntactically correct terms and, thus, restricts the usage of plain English language. Again, there is currently no additional \emph{composite structure}. However, the requirements engineer might later add requirements to this artefact and cluster them according to a composite structure. The requirements template is associated with a specific interpretation in the semantic viewpoint, yielding the pragmatics of the artefact. This interpretation could now be expressed in predicate logic. This means that the syntactic construct is associated with a logical formula that describes how the syntax is interpreted. Hence, predicate logic forms a domain in the semantic context. Considering this logical formula, we see that there is still room for interpretation. For example, it is not clear what the predicate \emph{``applied (brake)''} exactly stands for (pressing the pedal, electronic activation, etc.). This is still subject to interpretation based on the individual knowledge including experiences, expertise, and expectations, all contributing to the formation of a mental model. The role that individual knowledge plays in the interpretation of a structure gets smaller the more formal the used syntax and its semantic formation is defined (cf.\ mathematical expressions). 
\begin{figure*}[htbp]
\centering
\includegraphics[width=\textwidth]{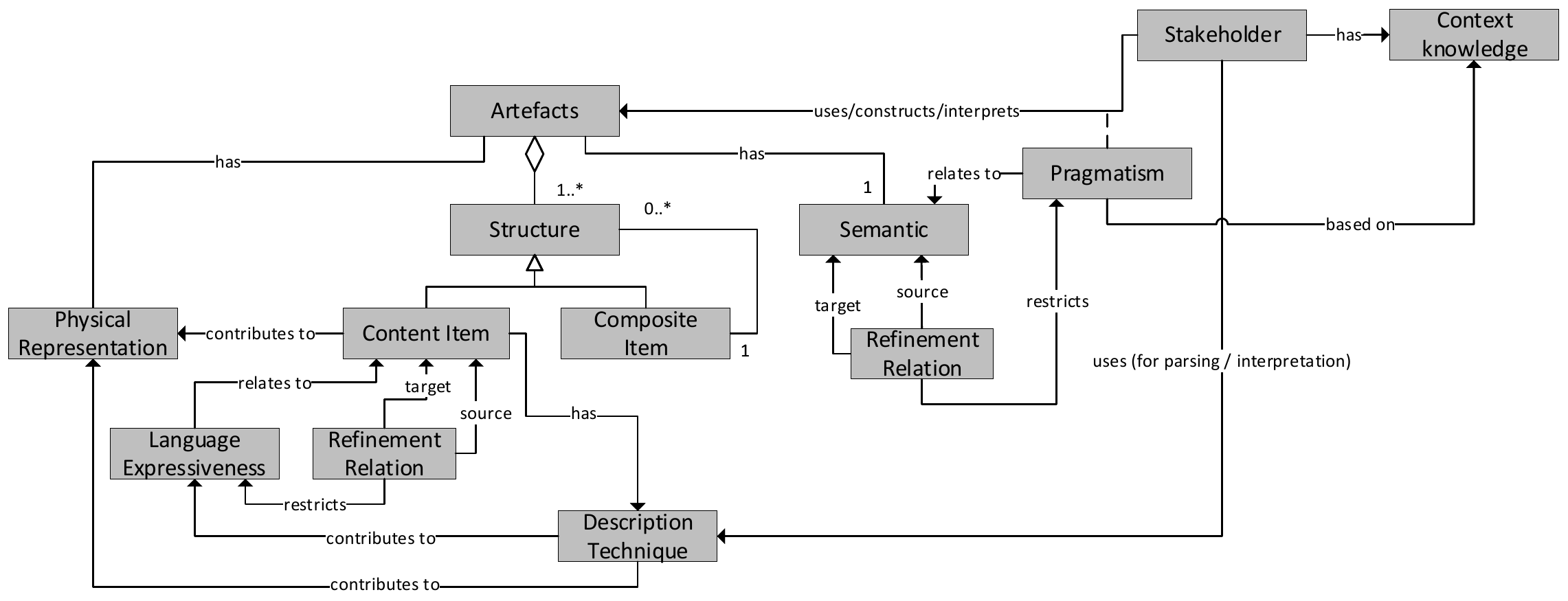}
\caption{Meta model for artefacts.}
\label{fig:metamodel}
\end{figure*}

By this small example, we already see that one piece of content (the meaning/intention) can be described via several (syntactic) structures, which in turn can each have several physical representations. The content is generally interpreted in a specific context of the stakeholder's individual knowledge. A summarising view on the concepts is given by the meta model presented in Fig.~\ref{fig:metamodel}. 

Finally, please note that an unambiguous description of an artefact's content using a formal syntax and semantics (e.g., formal logics) is possible to a certain degree, but there is a trade-off between precision and comprehensibility by humans of the used syntax. This trade-off strongly depends on the target group of an artefact, i.e. stakeholders, such as developers, who have to comprehend and interpret the artefact.

\section{Relating Artefacts: Refinement and Equivalence}
\label{sec:EquivArte}
As shown above, we have to consider three levels of perception, which are related by abstractions in the sense that there are multiple physical representations for one abstract syntactic structure and, likewise, multiple syntactic structures for one semantic content. However, these abstraction relations should not be mixed up with the distinction between an artefact instance and an artefact type (e.g., the artefact type \emph{``requirements specification''} vs.\ the artefact instance \emph{``requirements specification for Vehicle Braking System''}\/). This is also why we refrain from relating our levels of perception to the levels of abstraction as introduced along object orientation. When we speak about artefact types, we also need to speak about structure (syntax) types, whereas, when we speak about artefact instances, we also need to speak about instances of structures.

\begin{figure}[htbp]
\centering
\includegraphics[width=0.5\textwidth]{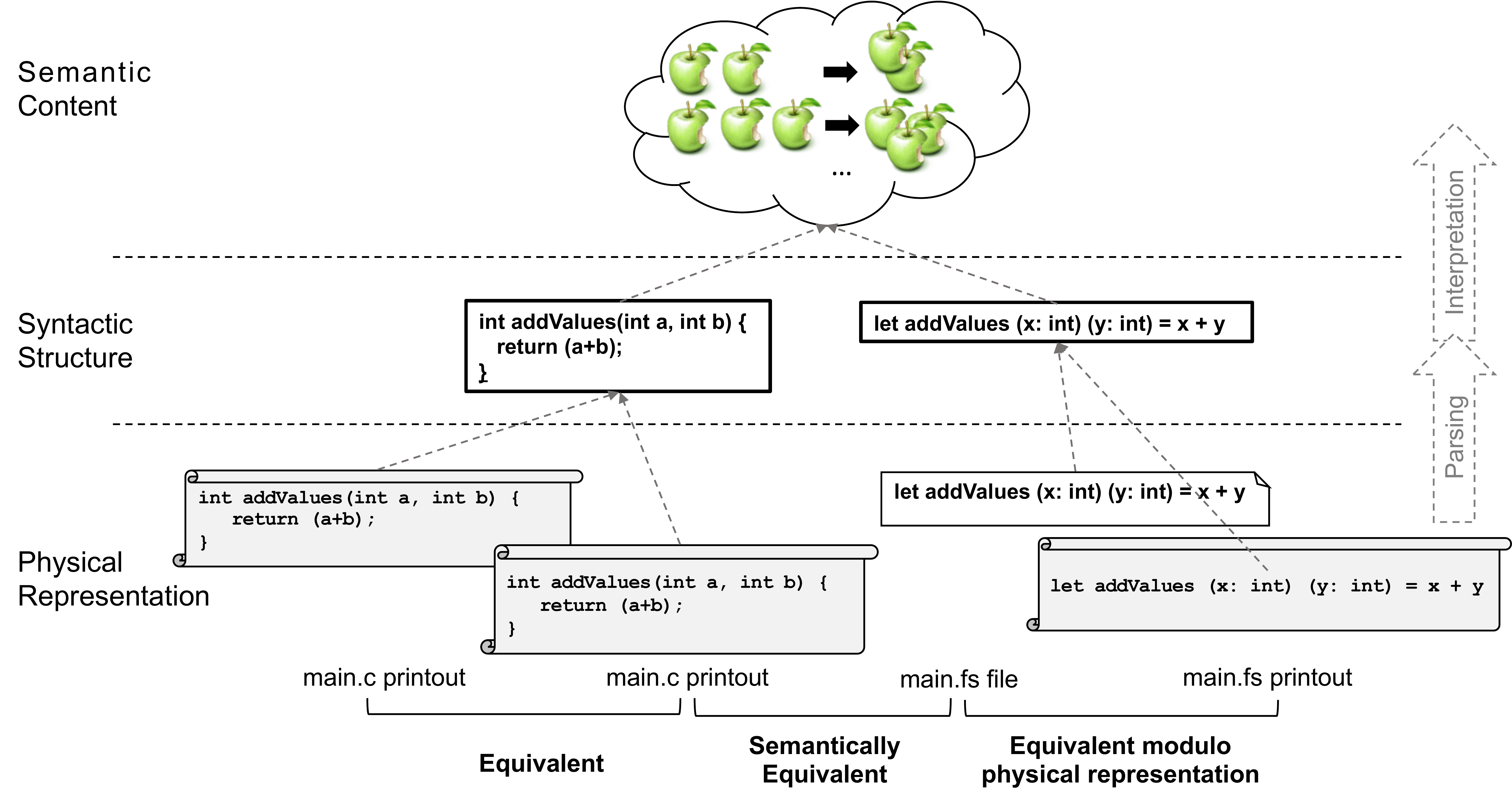}
\caption{Equivalent artefacts.}
\label{fig:equivalence}
\end{figure}

We now use our introduced abstraction relations (Fig.~\ref{fig:LevelsPerfecption}) to discuss the notion of \emph{artefact refinement and artefact equivalence}: To this end, we will take a closer look at the relations between artefacts and artefact elements at the different levels of perception by virtue of the example illustrated in Fig.~\ref{fig:equivalence}. Let us consider two programs, both incrementing a counter (variable) by one. The first program, called \texttt{main.c}, which is printed out twice, defines a simple C-function \texttt{addValues(int, int)}, while the second program, stored electronically as the file \texttt{main.fs} and printed out once, specifies the same function, but in the language F\#. Accordingly, we have four different artefacts. The two printouts of the first program are considered \emph{equivalent} in the narrowest sense, since they are equivalent at all levels of perception. The electronically stored file \texttt{main.fs} and its printout are \emph{equivalent modulo physical representation} since equality is restricted to the semantic content and syntactic structure. Finally, the printout of the first program and the electronically stored \texttt{main.fs} file (as well as its printout) are \emph{semantically equivalent}, i.e. their equivalence is restricted to the semantics expressed in the artefact, because the semantics of the two programming languages C and F\# leave no room for a meaningful different pragmatic interpretation. That is, both functions do exactly the same (modulo its ``syntactic sugar'').

According to these definitions, equivalence is a gradual concept ranging from an artefact's content to the physical representation, with each level of equivalence also requiring that the artefacts are equivalent at any level above. Consequently, the equivalence notions all require at least semantic equivalence. Conversely, any artefacts with different contents are not equivalent, regardless of the class of structure or the physical representation used. 

As we have seen in the small requirements example in section 3, the more rigid we are in the description of the syntactic structure and the semantic content, the less freedom is left in the interpretation of the content. By using a template for the requirements specification, we restrict the usage of the English language and by associating the template with a logical formula describing how to interpret the syntactic structure in predicate logic, we restrict the pragmatism by adding an additional domain in the semantic context. We say that an artefact is refined by restricting the language expressiveness and/or by adding additional domains in the semantic context, this way restricting the role individual knowledge plays in the semantic interpretation. Hence, the artefacts describing the right side of Fig.~\ref{fig:Example} are refinements of the artefacts on the left side. However, as pragmatics is inherently subjective for human interpretation, to find correspondence between such interpretations is still a difficult task (see, e.g.,~\cite{Kathleen+92} for an overview of representation and comparison techniques). In fact, often we have to compare and relate artefacts and their contents that are on different levels of formalisation and precision, such as informal in contrast to formal artefact contents (for details, see~\cite{Broy2018}). Then refinement is different from refinement notions as used in the field of formal methods. There, refinement is just logical implication. Going from informal to formal description is rather formalisation making informal content precise.

The notion of \emph{refined or equivalent artefacts} is useful in practice. For example, requirement and test artefacts can, in some cases, be understood as semantically equivalent. A requirement that is expressed as a sequence of interactions between the system and its environment (for instance, specified as a sequence diagram) and the corresponding test case that specifies consecutive stimulations of a system and expected results by a test script both describe the same content. However, the structures and the physical representations differ. The notion of semantic equivalence helps understanding the semantic relations between software engineering artefacts. From these relations, we can derive consistency rules and verification obligations. Moreover, they form the basis for tracing between artefacts.

\section{Related Work}
\label{sec:RelatedWork}
The paradigm of artefact orientation has gained a lot of attention during the last years, especially in requirements engineering as the problems of neglecting artefacts and their dependencies were recognised. In~\cite{MPKB10}, we provided a meta model for artefact-oriented requirements engineering where artefacts define structure and content of domain-specific requirements engineering results. Our work concluded with the following definition of a requirements engineering artefact: 

\begin{definition}
\label{def:ArtDefiOLD}
An artefact is a work product that is produced, modified, or used by a sequence of tasks that have value to a role. Artefacts are subject to quality assurance and version control and have specific types. They are hierarchically structured into content items that define single areas of responsibility and that are the output of a single task. Each content item encompasses at its lowest level of decomposition:
\begin{enumerate}
	\item Concepts: a concept defines the elements and their dependencies of domain specific description techniques used to represent the concern of a content item. Concepts have a specific type and can be decomposed to concept items. The latter differentiation is made if different items of a concept can be described with different techniques.
	\item Syntax: the syntax defines a concrete language or representation that can be chosen for a specific concept.
	\item Method: The method (or task) describes the sequence of steps that is performed to make use of a concept.
\end{enumerate}
\end{definition}
One exemplary instantiation of the meta model forms REMsES~\cite{Braun+10}, which provides a process guide for supporting requirements engineering processes in the automotive industry. There, the artefact model provides a basic structure for the definition of the artefacts, their assignment to abstraction levels and content categories, and the relations between the artefacts. It further defines general control flow dependencies within requirements engineering processes. Further fields of applications can be found in the broader field of software engineering process modelling, e.g. for synchronising various software process models via artefacts as interfaces~\cite{KMG13} or for modelling artefact-focused method-neutral software process improvement models~\cite{Kuhrmann+15}.

Another view on the term ``artefact'' is provided by Silva et al.~\cite{Silva_softwareartifact} who structure artefacts by defining a meta model that explicitly lists specifying the basic elements (fragments) to create generic software artefacts. The fragments consist of simple types such as text and images, complex types which use simple types to define structured information such as diagrams and tables, as well as specific types to extend the model for process specific structures. In their model, artefacts that are developed in a given software development process are defined as an instance of an artefact meta class which adds process specific semantic that is required to specify the artefacts present in a given software development process. Their artefact meta class defines more than 100 classes to organise artefacts among sections, subsections and other types of information. As the basic elements are derived empirically by examining software processes based on the Software and System Engineering Process Meta-model (SPEM~\cite{SPEM}), such as RUP, but also SCRUM, their approach is also process-dependent as in our own previous work.

Another example stems from an area known as method engineering~\cite{Braun+05} which emerged in response to shortcomings also reported by Parnas and Clements~\cite{PC86} including that processes are too often performed mindlessly or even faked without awareness of the reasons why a process step should (or should not) be executed, and without awareness of how to structure and specify the results. Consequently, artefacts were often not reproducible, not measurable, incomplete, and inconsistent with no clear terminology~\cite{MWLBC10}. The underlying philosophy of method engineering defines a situation-specific process by a set of methods to be performed in a particular order by a specific set of roles. Although this research area fostered the discussion on adaptable processes, it still does not cope with the various challenges of today's project environments. Moreover, there is still no common agreement on the structure and semantics of artefact-based methodologies \cite{MPKB10}. As available approaches (e.g., SPEM-related approaches) further focus on methods and description techniques as well as the complex dependencies between the methods rather than on clear result structures, contents, and dependencies among the artefacts, project participants remain unaware of how to create consistent artefacts in their projects independently of underlying restrictive processes~\cite{Kuhrmann+15}. 

As discussed in the introductory part of the paper at hands, our previous definition stems from a process-centric view, same as those of the examples given above, and the definition is thus valid for a process-dependent context only as processes (being a series of tasks) are used as a means for the definition. Secondly, and more importantly, we did not consider at all the different ways to provide semantics of the artefacts. This makes it inherently difficult to transfer the concept to fields that are not dependent on specific processes (e.g. the general area of model-based development).

Overall, we can observe that different research areas came up with their own definition of what constitutes an artefact in order to examine or transform it at certain levels of perception. However, not all definitions are explicit and relationship between the levels of perception often remain vague. For instance, model versioning and comparison~\cite{Altmanninger+09} consider a model as an artefact; unclear is if they refer to a physical file, a graph according to a certain meta model or to content representing an abstraction of some part of reality. Moreover, those levels of perception are referred to by different authors in the same research area by different names, e.g., static instead of syntactic~\cite{Nejati+07}. The lack of a common vocabulary and precise understanding limits the comparability of different techniques and approaches, as identified independently in~\cite{Brunet+06} Closing this gap of a common vocabulary was in scope of our paper at hands.

In a recent paper, Broy~\cite{Broy2018} distinguishes for an artefact between logical content, content representation (i.e. its syntax), and its physical representation, and defines the notion of content chunks. Content chunks are either elementary parts of an artefact, or a set of content chunks. The meaning of a content chunk is assumed to be represented by logic, which defines a semantic theory defining the meaning of the content chunk. Based on this definition, he introduces formal links and traces between requirements, functional specifications, and architectures relating the content of the respective artefacts. Note that the meta model presented in [[Figure 3]] is in tune with this approach. Hence, Broy's work constitutes one application on basis of our notion of artefacts and their relationships discussed in the manuscript at hands.

\section{Conclusion}
\label{sec:Conclusion}
In this paper, we critically reflected on the various interpretations of the term ``artefact'' in context of software engineering and discussed its basic concepts by taking different views independent of potentially surrounding processes or related concepts. Our discussion leads to a revised definition of the notion of an artefact in the context of Software Engineering:

\begin{definition}
\label{def:ArtNEW}
An artefact is a self-contained work result, having a context-specific purpose and constituting a physical representation, a syntactic structure and a semantic content, forming three levels of perception.
\end{definition}
The exact meaning of the semantic content, the pragmatics, depends on the parsing, which extracts the syntactic structure and the way in which a semantic content is finally interpreted in the individual context of the stakeholder. Hence, an artefact is uniquely defined by the three levels of perception which result from the application of a given parser and interpreter. A given process then can define which artefacts to be produced in the course of a software development project. By taking our descriptive view, however, we could characterise artefacts independent of their surrounding processes and, thus, provide a clear understanding of the main concepts, in turn supporting a clear notion of artefact orientation and those software engineering practices that rely on this notion.

Researchers and practitioners as well can now in fact relate their discussions and various contributions to our field in context of that term as already shown in~\cite{Broy2018}. By introducing the notion of refinement and equivalence, we can now also better understand the relations between software engineering artefacts and derive, for instance, consistency rules. The notion of equivalence could be further extended to cover, for instance, settings with different variants and versions of artefacts in the future. 

With our paper at hands, we contribute to a common understanding and hope to support steering the development of artefact-centric contributions such as ones providing or relying on sets of interdependent software engineering artefacts, i.e. artefact models. Having a clear picture of the notion of artefacts and their relations allows us to embed this concept into the broader context of engineering practices including software process models or, more generally speaking, seamless engineering processes, which up to now was done based on domain-specific, process-dependent, and too often very narrow views. Modelling the relationships between artefacts is therefore a key for a better usage of models in system engineering. To achieve this however, further work is required to cover traceability in full generality where semantically richer concepts for and relationships between artefacts provide a rigorous basis for semantic linking and tracing within and between artefacts~\cite{Broy2018}.

\bibliographystyle{plain}
\bibliography{Artefact}

\end{document}